\documentclass[aip,apl,twocolumn,a4paper,10pt]{revtex4}
\usepackage{color}
\usepackage{epsfig}
\usepackage{amssymb}
\usepackage[dvipdfmx]{hyperref}
\hypersetup{colorlinks=true,citecolor=blue,linkcolor=blue,urlcolor=black}
\usepackage[all]{hypcap}
\usepackage{url}

\begin{document}

\title{The Influence of Silicon Nanoclusters on the Optical Properties of a-SiN$_x$ Samples: A Theoretical Study }

\author{\firstname{Roberto} \surname{Guerra}}
\affiliation{Dipartimento di Scienze e Metodi dell'Ingegneria and Centro Interdipartimentale En\&Tech, Universit\`a di Modena e Reggio Emilia, via Amendola 2 Pad.\ Morselli, I-42122 Reggio Emilia, Italy.}

\author{\firstname{Mariella} \surname{Ippolito}}
\affiliation{Consorzio Interuniversitario per le Applicazioni di Supercalcolo Per Universit\`a e Ricerca (CASPUR), Via dei Tizii 6, I-00185 Roma, Italy.}

\author{\firstname{Simone} \surname{Meloni}}
\affiliation{Consorzio Interuniversitario per le Applicazioni di Supercalcolo Per Universit\`a e Ricerca (CASPUR), Via dei Tizii 6, I-00185 Roma, Italy.}
\affiliation{School of Physics, Room 302 UCD-EMSC, University College Dublin, Belfield, Dublin 4, Ireland.}

\author{\firstname{Stefano} \surname{Ossicini}}
\affiliation{Dipartimento di Scienze e Metodi dell'Ingegneria and Centro Interdipartimentale En\&Tech, Universit\`a di Modena e Reggio Emilia, via Amendola 2 Pad.\ Morselli, I-42122 Reggio Emilia, Italy.}

\begin{abstract}
\noindent By means of {\it ab-initio} calculations we investigate the optical properties of pure a-SiN$_x$ samples, with $x \in [0.4, 1.8]$, and samples embedding silicon nanoclusters (NCs) of diameter $0.5$$\leq$$d$$\leq$$1.0$~nm. In the pure samples the optical absorption gap and the radiative recombination rate vary according to the concentration of Si-N bonds. In the presence of NCs the radiative rate of the samples is barely affected, indicating that the intense photoluminescence of experimental samples is mostly due to the matrix itself rather than to the NCs. Besides, we evidence an important role of Si-N-Si bonds at the NC/matrix interface in the observed photoluminescence trend.
\end{abstract}

\maketitle

Over the last few years silicon nitride alloys (a-SiN$_x$) emerged as possible alternatives to silicon oxides (a-SiO$_x$) in silicon-based optoelectronic devices. What makes SiN$_x$  appealing is its larger static and optical dielectric constant with respect to SiO$_2$ ($8.19$ and $4.33$ for a-Si$_3$N$_4$, $4.6$ and $2.46$ for a-SiO$_2$, respectively), which prevents the leakage of tunneling currents in nanostructured devices. Also very important is its higher stability at large electric fields and high temperatures (commercial a-Si$_3$N$_4$ samples melt at about 2175 K compared to about 2000 K for a-SiO$_2$). Moreover, the reduced band gap of a-SiN$_x$ with respect to SiO$_x$ ($4.7$ eV for a-Si$_3$N$_4$ and $8.9$ eV for a-SiO$_2$) requires smaller operating voltages to obtain the same electron and hole currents. In analogy to the case of a-SiO$_2$,\cite{revlockwood} the photoluminescence (PL) and electroluminescence (EL) of a-SiN$_x$ was attributed to the presence of Si nanoclusters (NCs).\cite{park,mercaldo,cinesi1,cinesi2} However, recent experiments\cite{kistner} indicate, consistently with hypothesis formulated previously,\cite{dashpande} that a-SiN$_x$ is photoluminescent on its own, i.e.\ even when Si-NCs are not present in the sample. Clarifying this point is a prerequisite for further developments of this material. In fact, a typical strategy for tuning the optical properties of materials embedding NCs is to control the size, shape, density, and other structural properties of the nanoparticles. Of course, this strategy would be ineffective if the PL of a-SiN$_x$ is not due to embedded NCs.  
Establishing the effect of Si-NCs on the optical properties of a-SiN$_x$ samples is also important in view of applications of this material in others fields, such as light-emitting diodes,\cite{park} lasers,\cite{monroy} and solar cells.\cite{adachi}
\\In this letter we try to shed some light on this point by computing the optical gap and the finite temperature radiative recombination rate (RR) of pure samples (i.e.\ not containing NCs) at various stoichiometries, and of a-SiN$_x$ samples containing Si-NCs of diameter $0.5$~nm $\leq d \leq 1.0$~nm. This will also allow us to estimate the effect of the size of the NCs on the optical properties of the material.
\\Let us start the description of our results from the pure a-SiN$_x$ systems. Samples of a-SiN$_x$ of stoichiometry $0.4 \leq x \leq 1.8$ containing as many as 224 atoms are prepared following a very careful quenching-from-the-melt procedure, requiring long simulation times of about 1~ns. The simulations were performed using the classical MD package {\small CMPTool} \cite{CMPTool} and the atomic interactions are governed by the Billeter et al. force field.\cite{Billeter} More details about the simulation protocol are reported in Ref.\ [\onlinecite{mariella}], where it was shown that the adopted procedure produces samples that are consistent with the experimental ones in terms of concentration of various types of defects and bond lengths. Typically, the finite temperature configurations produced by this approach are highly stressed, resulting in an electronic structure that is difficult to analyze (e.g.\ it contains a large number of gap states). This is not an artifact of the method as in real conditions these configurations are averaged over a suitable statistical ensemble. However, the calculation of the ensemble averaged spectrum of samples of this size is unaffordable. We take, therefore, a different approach and compute the spectrum over the equilibrium configuration corresponding to one finite temperature configuration sampled along the classical MD simulation. This equilibrium configuration is obtained by optimizing the geometry of the systems using the real-space DFT package {\small SIESTA}.\cite{siesta} Electronic and optical properties of the relaxed structures are then obtained by reciprocal-space DFT calculations using the {\small Quantum ESPRESSO} package.\cite{espresso} These latter calculations are performed using norm-conserving pseudopotentials within the local-density approximation (LDA). An energy cutoff of 60 Ry on the plane-wave basis set is imposed, and a gaussian smearing function of $0.25$ eV is applied on the electronic population in order to circumvent convergency issues related to the presence of dangling bonds.
\begin{figure}[b!]
 \centering
 \psfig{file=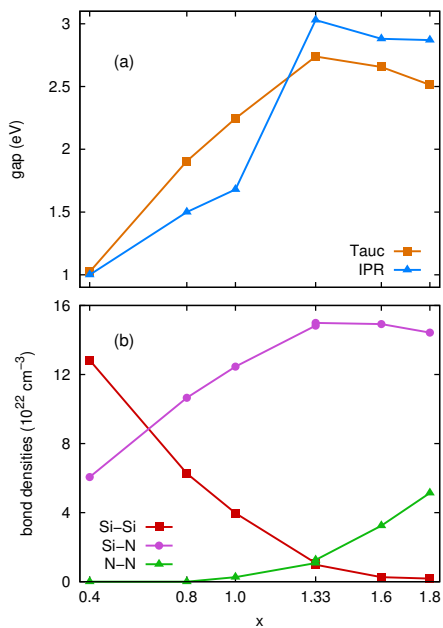,width=\columnwidth,angle=270}
 \caption{\it (a) a-SiN$_x$ gap vs $x$ curve estimated by using the Tauc (squares) and IPR (triangles) methods. (b) Si-Si, Si-N, and N-N bond densities, as a function of the stoichiometry of the sample. }\label{Fig1}
\end{figure}
\\The determination of the gap, not an easy task in disordered systems,\cite{sasaki,guraya,davis} is performed using two methods:  the Tauc method,\cite{tauc} frequently used with both experimental and computational data, and the Inverse Participation Ratio (IPR) method.\cite{justo} 
The Tauc approach assumes that at the absorption onset holds the relation $\hbar\omega N(\hbar\omega) \propto (\hbar\omega - E_g)^2$, where $N(\hbar\omega)$ is the number of interband transitions at a given energy, i.e.\ the spectral density, and $E_g$ is the optical gap. $E_g$ can, therefore, be obtained by the fitting of the above relation.\cite{valladares_prb}
The IPR measures the degree of localization of an orbital. IPR~=~1 means a maximally dispersed orbital, IPR~=~$\infty$ a maximally localized one. Assuming that highly localized states (defect states) do not contribute to the spectrum, we can estimate the gap by considering only those states with an IPR below some predetermined threshold.\cite{justo} In Fig.~\ref{Fig1}a we report the values of the gap, calculated using both methods, as a function of the composition of the a-SiN$_x$ sample. The gaps reported here are in nice agreement with experimental observation on hydrogen-free samples,\cite{davis,sasaki} beside a systematic underestimation of the calculated gaps of at most $0.5$~eV that is most likely due to the well known limitation of DFT in describing excited states.\cite{payne} In addition,  Both methods show that $E_g(x)$ increases with $x$ up to the stoichiometric composition and then gently decreases. Following Manca,\cite{manca} this trend can be explained by noticing that at $x=1.33$ the density of the more energetic Si-N bonds is maximum, while in sub and supra-soichiometric samples their density decreases in favor of less energetic Si-Si and N-N bonds (see Fig.~\ref{Fig1}b).
The IPR gap shows a sudden increase in the range $x = 0.8-1.0$, which is consistent with experimental observations.\cite{davis,guraya} On the contrary, the gap curve obtained by the Tauc method has an unexpected convex profile over the entire $x$ range. This is because the latter method does not take into account the orbitals' localization. Despite the above limitation, the Tauc method gives a gap with a correct overall trend, consistent with experimental and IPR results.   
\begin{figure}[b!]
 \centering
 \psfig{file=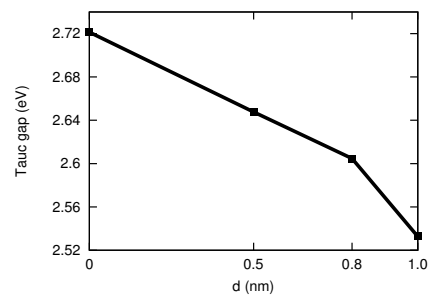,width=4cm,angle=270}
 \caption{\it Tauc gap as a function of the size of the embedded Si-NCs. The gap value of the pure a-SiN$_{1.33}$ matrix is also reported and denoted by $d = 0$~nm.
 }\label{Fig2}
\end{figure}

\begin{figure}[b!]
 \centering
 \psfig{file=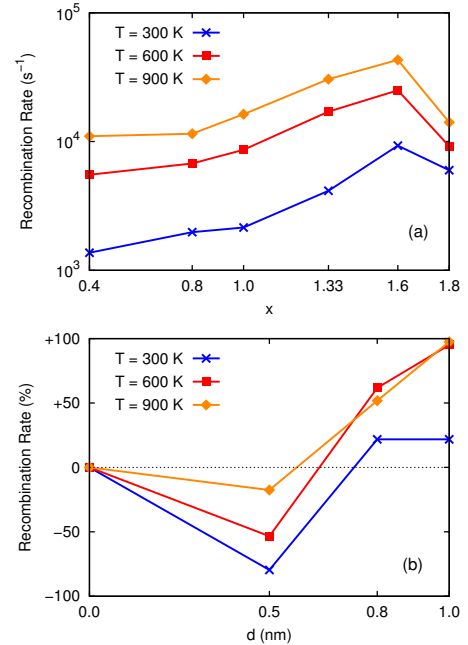,width=\columnwidth,angle=270}
 \caption{\it (a) Thermally averaged RR for pure a-SiN$_x$ samples as a function of their stoichiometry; (b) percentage variation of the RR with respect to the pure a-Si$_3$N$_4$ matrix, as a function of the NC diameter $d$ ($d = 0$~nm denotes the pure matrix). }\label{Fig3}
\end{figure}

Moving to the effect of Si-NCs on the gap, we studied samples containing Si-NCs of three diameters: $d = 0.5, 0.8$ and $1.0$~nm.  These samples are prepared by placing a spherical Si-NC in a cavity of suitable size created in an a-Si$_3$N$_4$ sample. The a-Si$_3$N$_4$ sample used here is larger than those considered above and contains as many as 882 atoms. Such a large sample is necessary in order to avoid the interaction of the nanoparticle with its periodic images and so that the bulk structure of the matrix is recovered within the boundary of the simulation box. The so prepared samples are relaxed by a 50~ps long classical MD simulations and then further relaxed by {\itshape ab-initio} geometry optimizations.
At a variance with the pure matrix case, on samples embedding small NCs the IPR method cannot be applied. In fact the IPR removes all the NC confined states that may contribute to the optical absorption. At the same time, no low-IPR (bulk-like) states are present in the matrix gap for NCs of this size. As a result, the $E_g$ of systems embedding NCs is the same as that of the corresponding pure matrices. On the contrary, the Tauc method, which is based on the fitting of the spectral density over a large energy domain, keeps track of the NC's states lying in the valence and conduction bands. Thus, in the latter case the resulting $E_g(d)$ curve presents a regular trend with the size of the nanoparticle (see Fig.~\ref{Fig2}). As expected, the $E_g$ decreases with the diameter of the NC. The modest variation of the gap with the NC size is likely determined by the presence of defects, that are known to cause a weakening of quantum size effects.\cite{nguyen2012} Also, the convex shape of the trend is in contrast with the $d^{-\alpha}$ trend expected for systems governed by quantum confinement (QC).\cite{park} By comparing our results with those of Park et al.\cite{park}, we suggest that the QC-like behavior is recovered for NCs of diameter $\gtrsim 1$~nm.
\\Let us move to the study of the optical efficiency with the stoichiometry of the pure samples and with the size of the Si-NCs. The optical efficiency depends on the (thermally averaged) RR, that we compute according to the method described in Ref. [\onlinecite{PRB3}]. It is worth stressing that in presence of defective samples the RR may critically depends on the defect type and concentration, rather than the bulk optical properties. As a result, in highly defective samples the defect sites may govern the total RR. It is therefore important to consider samples obtained following the same procedure in order to have a meaningful comparison of the RR. The RR at 300, 600, and 900~K for the pure a-SiN$_x$ samples and the a-Si$_3$N$_4$ containing Si-NCs are shown in Fig.~\ref{Fig3}a and \ref{Fig3}b, respectively. For pure a-SiN$_x$ samples, at high $T$ the RR increases with the nitrogen content up to $x$=$1.6$ and then decreases. This is consistent with recent experimental observations.\cite{hiller} The larger recombination times ($\sim$20~$\mu$s at least) with respect to experimental $\mu$s--ns-like PL decay times,\cite{negro} are mostly referable to the lack of H atoms and of the consequent hydrogenic passivation in our samples.\cite{wilson,nguyen2012} In addition, other approximations used in our calculations, such as the LDA, the value of the static dielectric constant, the lack of phonon interaction and non-radiative decay channels, indicate that the calculated RR should be considered a lower bound. Also, the amorphous nature of small NCs is a known quenching factor of their optical activity.\cite{negro} It is worth stressing that despite the systematic underestimation of the computed RR, its dependence on the matrix stoichiometry and NC size should not be affected by the above approximations.
\\For samples embedding NCs, the RR do not present a clear trend with the size of the nanoparticle. By taking the $T$=600 curve as reference, we observe a RR reduction ($d$=0.5) or increase ($d$=0.8,1.0) of up to a factor $2$ with respect to the pure matrix. Such variation of the RR is probably connected to the emergence of different kind and number of defects at the Si/Si$_3$N$_4$ interface, rather than to the presence of the NC itself. This explanation is also compatible with recent observations that hypothesize a secondary role of NC in the PL emission spectra of samples formed by varying the silicon excess (in our case $x$=1.23 at $d$=1.0~nm).\cite{kistner,sahu} 
\\Finally, the present result supports the Dal Negro et al.\cite{negro} hypothesis that the PL of a-SiN$_x$ samples is due to states associated to Si-N-Si bridge bonds at the NC/matrix interface. In fact, the number of such bonds, and the associated states, increases with the size of the nanoparticle and the PL intensity grows accordingly, at least in a limited range of $d$.
\\The small dependency of the PL on the NC size seems to indicate that hydrogen-free samples containing NCs are less suitable for producing Si-based emitters\cite{kistner}, while the possibility of tuning the optical absorption energy might be exploited in forthcoming photovoltaic applications.\cite{mercaldo}

\vspace{0.25cm}
\noindent \footnotesize{ Computational resources were made available by CINECA-ISCRA parallel computing initiative and by CASPUR. R.G. and S.O. acknowledge financial support from the European Community's Seventh Framework Programme (FP7/2007-2013) under Grant No. 245977. S.M. acknowledges financial support from the SFI under the SFI-PI grant number 08-IN.1-I1869 and the European Community under the Marie Curie Intra-European Fellowship for Career Development grant number 255406. }

\end{document}